\documentclass[oneside,english]{autart}
\usepackage[T1]{fontenc}
\usepackage[latin9]{inputenc}
\usepackage{color}
\usepackage{amsmath}
\usepackage{amssymb}
\usepackage{graphicx}

\makeatletter
\makeatletter
\def\old@comma{,}
\catcode`\,=13
\def,{%
  		\ifmmode%
    		\old@comma\discretionary{}{}{}%
  		\else%
    		\old@comma%
  		\fi%
}
\makeatother
\usepackage{cite}

\makeatother

\usepackage{babel}
\begin{document}
\begin{frontmatter}

\title{Decentralized Rendezvous of Nonholonomic Robots with \textcolor{black}{Sensing
and }Connectivity Constraints}

\author{}\author[ufl]{Zhen Kan}\ead{kanzhen0322@ufl.edu},
\author[ufl]{Justin Klotz}\ead{jklotz@ufl.edu},
\author[reef]{Eduardo L. Pasiliao Jr}\ead{pasiliao@eglin.af.mil},
\author[ufl2]{John M. Shea}\ead{jshea@ece.ufl.edu},
\author[ufl]{Warren E. Dixon}\ead{wdixon@ufl.edu},
\address[ufl]{Department of Mechanical and Aerospace Engineering, University of Florida, Gainesville, USA}
\address[reef]{Munitions Directorate, Air Force Research Laboratory, Eglin AFB, FL 32542, USA.}
\address[ufl2]{Department of Electrical and Computer Engineering, University of Florida, Gainesville, USA}
\thanks{This research is supported in part by NSF award numbers 1161260, 1217908, and a contract with the AFRL Mathematical Modeling and Optimization Institute. Any opinions, findings and conclusions or recommendations expressed in this material are those of the authors and do not necessarily reflect the views of the sponsoring agency. }
\begin{abstract}
\textcolor{black}{A group of wheeled robots with nonholonomic constraints
is considered to rendezvous at a common specified setpoint with a
desired orientation while maintaining network connectivity and ensuring
collision avoidance within the robots. Given communication and sensing
constraints for each robot, only a subset of the robots are aware
or informed of the global destination, and the remaining robots must
move within the network connectivity constraint so that the informed
robots can guide the group to the goal. The mobile robots are also
required to avoid collisions with each other outside a neighborhood
of the common rendezvous point. To achieve the rendezvous control
objective, decentralized time-varying controllers are developed based
on a navigation function framework to steer the robots to perform
rendezvous while preserving network connectivity and ensuring collision
avoidance. Only local sensing feedback, which includes position feedback
from immediate neighbors and absolute orientation measurement, is
used to navigate the robots and enables radio silence during navigation.
Simulation results demonstrate the performance of the developed approach. }
\end{abstract}
\end{frontmatter}

\section{\textcolor{black}{Introduction}}

\textcolor{black}{Distributed cooperative control of networked multi-agent
systems has attracted considerable interest. One particular cooperative
control problem is the rendezvous problem, where a number of agents
arrive at a predefined destination simultaneously, ideally using limited
information from the environment and team members. Some example applications
of the rendezvous problem are cooperative strike and cooperative jamming
in \cite{McLain2005} and \cite{Tekin2008}. In the cooperative strike
scenario, multiple strikes are executed on a target simultaneously
by firing from different locations. In cooperative jamming of a wireless
communication network with eavesdroppers, noisy signals are transmitted
to jam the eavesdroppers at the same time when the source transmits
the message signal. Spacecraft docking, air-to-air refueling, and
the interception of an incoming missile can also be considered as
rendezvous problems. In these applications, coordination and collaboration
are crucial to performance, and agents are required to communicate
and coordinate their movements with others to achieve rendezvous.}

\textcolor{black}{Several rendezvous results are reported in \cite{Lin2004,Lin2007,Lin2007a}.
Convergence to a common point for a group of autonomous mobile agents
is studied in \cite{Lin2004}. In \cite{Lin2007} and \cite{Lin2007a},
synchronized and unsynchronized strategies are developed to drive
mobile agents to a single unspecified location by using only position
feedback from its sensing regions. A common assumption in \cite{Lin2004,Lin2007,Lin2007a}
is that the network remains connected during the motion evolution,
allowing constant interaction between agents. However, the assumption
of network connectivity is not always practical. Typically, each agent
can only make decisions based on the local information from immediate
neighbors within a certain region due to sensing and communication
constraints. Since communication/sensing links generally depend on
the distance between agents, agent motion may cause the underlying
network to disconnect. If the network disconnects, certain agents
may no longer be able to communicate and coordinate their motion,
leading to a failure of cooperative tasks.}

\textcolor{black}{Recent results such as \cite{Cort'es2006,dimarogonas2007,Su2010a,Gustavi2010,Xiao.Wang.ea2012,Hui2011}
have focused on maintaining network connectivity when performing rendezvous
tasks. A circumcenter algorithm is proposed in \cite{Cort'es2006}
to avoid the loss of existing links between agents. In \cite{dimarogonas2007,Su2010a,Gustavi2010}
a potential field-based distributed approach is developed to prevent
partitioning in the underlying graph by using local information from
each agent's immediate neighbors. The results in \cite{Xiao.Wang.ea2012}
provide a connectivity-preserving protocol for rendezvous of a discrete-time
multi-agent system, and a hybrid dynamic rendezvous protocol is designed
in \cite{Hui2011} to address finite-time rendezvous problems while
preserving network connectivity. However, most of the aforementioned
works only consider linear motion models. Although agents with nonholonomic
kinematics are considered in \cite{dimarogonas2007}, like other results
such as \cite{Lin2004,Lin2007,Lin2007a,Hui2011}, the agents can only
converge to a destination determined by the initial deployment. A
dipolar navigation function was proposed and a discontinuous time-invariant
controller was developed for a multi-robot system in \cite{loizou2008}
to perform nonholonomic navigation for networked robots. The dipolar
navigation function is a particular class of potential functions,
which is developed from \cite{Rimon_1990} and \cite{Rimon1992} such
that the negative gradient field does not have local minima, and the
closed-loop navigation function guarantees convergence to the global
minimum. The result in \cite{loizou2008} was then extended to navigate
a nonholonomic system in three dimensions in \cite{Roussos2008}.
Other recent results focused on nonholonomic systems with various
cooperative tasks such as formation control and flocking are reported
in\cite{Liu2012,Dong2008,Dong2009,Consolini2008}. However, network
connectivity is not considered in \cite{loizou2008,Roussos2008,Liu2012,Dong2008,Dong2009,Consolini2008}.}

\textcolor{black}{The rendezvous problem for mobile robots with nonholonomic
constraints is studied in this work, and the objective is to reach
a common specified setpoint with a desired orientation. Only a small
subset of robots (i.e., informed agents) are assumed to be equipped
with advanced sensors (e.g., GPS) and provided with global knowledge
of the destination, while the remaining robots (i.e., followers) only
have a range sensor (e.g., a passive range sensor such as a camera,
or active sensors such as sonar, laser, or radar), which provides
local feedback of the relative trajectory of other robots within a
limited sensing region. Since the follower robots are not aware of
the global position of the destination, they have to stay connected
with the informed agents when performing rendezvous. To avoid collision
among robots, the workspace is divided into a collision-free region
and a rendezvous region. Particularly, the robots are required to
avoid collisions with other robots outside a neighborhood of the common
goal. Based on our preliminary efforts in \cite{Kan2011,Kan.Klotz.ea2012,Kan.Dani.ea2012},
a decentralized time-varying controller, using only local sensing
feedback from its immediate neighbors, is designed to stabilize the
robots at the specified destination while preserving network connectivity
and ensuring collision avoidance. The developed decentralized controller
only uses local sensing information and no inter-agent communication
is required (i.e., communication-free global decentralized group behavior).
Although network connectivity is maintained so that radio communication
is available when required for various tasks, communication is not
required for navigation. Using the navigation function framework,
the multi-robot system is guaranteed to rendezvous at a common destination
with a desired orientation without being trapped by local minima from
almost all initial conditions, excluding a set of measure zero. Compared
to \cite{Kan.Dani.ea2012} where the formation control for a group
of agents with fully actuated dynamics is investigated, networked
mobile robots with nonholonomic constraints are considered in this
work. Unlike our centralized result in \cite{Kan2011} or our preliminary
result in \cite{Kan.Klotz.ea2012} in which all the robots are required
to know the goal destination and only undirected interaction between
robots are considered, the current result models the interaction among
robots as a digraph, and only requires a subset of the robots (i.e.,
one or more) to have knowledge of the global position of the destination
and the desired orientation. This advancement reduces required resources
and sensor loads on the remaining robots. Within this setting, the
informed subset of robots can perform a task-level controller, while
the remaining robots just execute a local interaction-based strategy.
Moreover, the developed controller allows the robots to rendezvous
at any desired destination, versus an unspecified destination determined
by their initial deployment as in \cite{Lin2004,Lin2007,Lin2007a,Hui2011,dimarogonas2007}.
The result can also be extended by replacing the objective function
in the navigation function to accommodate different tasks, such as
formation control, flocking, and other applications.}

\section{\textcolor{black}{Problem Formulation\label{sec:problem}}}

\textcolor{black}{Consider $N$ networked mobile robots operating
in a workspace $\mathcal{F}$, where $\mathcal{F}$ is a bounded disk
area with radius $R_{w}$. Each robot in $\mathcal{F}$ moves according
to the following nonholonomic kinematics:
\begin{equation}
\dot{q}_{i}=\left[\begin{array}{cc}
\cos\theta_{i} & 0\\
\sin\theta_{i} & 0\\
0 & 1
\end{array}\right]\left[\begin{array}{c}
v_{i}\left(t\right)\\
\omega_{i}\left(t\right)
\end{array}\right],\text{ }i=1,\cdots,N\label{eq:dynamics}
\end{equation}
where $q_{i}\left(t\right)\triangleq\left[\begin{array}{cc}
p_{i}^{T}\left(t\right) & \theta_{i}\left(t\right)\end{array}\right]^{T}\in\mathbb{R}^{3}$ denotes the states of robot $i,$ with $p_{i}\left(t\right)\triangleq\left[\begin{array}{cc}
x_{i}\left(t\right) & y_{i}\left(t\right)\end{array}\right]^{T}\in\mathbb{R}^{2}$ denoting the position of robot $i$, and $\theta_{i}\left(t\right)\in\left(-\pi,\pi\right]$
denoting the robot orientation with respect to the global coordinate
frame in $\mathcal{F}$. In (\ref{eq:dynamics}), $v_{i}\left(t\right),$
$\omega_{i}\left(t\right)\in\mathbb{R}$ are the control inputs that
represent the linear and angular velocity of robot $i,$ respectively.}

\textcolor{black}{The subsequent development is based on the assumption
that all robots have equal actuation capabilities and each robot has
sensing and communication limitations encoded by a disk area with
radius $R,$ which indicates that two moving robots can sense and
communicate with each other as long as they stay within a distance
of $R.$ We also assume that only a subset of the robots, called informed
robots, are provided with knowledge of the destination, while the
other robots can only use local state feedback (i.e., position feedback
from immediate neighbors and absolute orientation measurement). Furthermore,
while multiple informed robots may be used for rendezvous, the analysis
and results of this work are focused on a single informed robot. The
techniques proposed in this work could be extended to the case of
multiple informed robots by using containment control, as explained
in Remark 2. The interaction among the robots is modeled as a directed
graph $\mathcal{G}\left(t\right)=\left(\mathcal{V},\mathcal{E}(t)\right)$,
where the node set $\mathcal{V=}\left\{ 1,\cdots,N\right\} $ represents
the group of robots, and the edge set $\mathcal{E}(t)$ denotes time-varying
edges. The set of informed robots and followers are denoted as $\mathcal{V}_{L}$
and $\mathcal{V}_{F}$, respectively, such that $\mathcal{V}_{L}\cup\mathcal{V}_{F}=\mathcal{V}$
and $\mathcal{V}_{L}\cap\mathcal{V}_{F}=\emptyset$. Let $\mathcal{V}_{L}=\left\{ 1\right\} $
and $\mathcal{V}_{F}=\left\{ 2,\cdots,N\right\} $. A directed edge
$\left(i,j\right)\in$ $\mathcal{E}$ in $\mathcal{G}\left(t\right)$
exists between node $i$ and $j$ if their relative distance $d_{ij}\triangleq\left\Vert p_{i}-p_{j}\right\Vert \in\mathbb{R}^{+}$
is less than $R$. The directed edge $\left(i,j\right)$ indicates
that node $i$ is able to access the states (i.e., position and orientation)
of node $j$ through local sensing, but not vice versa$.$ Accordingly,
node $j$ is a neighbor of node $i$ (also called the parent of node
$i$), and the neighbor set of node $i$ is denoted as $\mathcal{N}_{i}=\left\{ j\text{ }|\text{ }\left(i,j\right)\in\mathcal{E}\right\} $,
which includes the nodes that can be sensed. A directed spanning tree
is a directed graph, where every node has one parent except for one
node, called the root, and the root node has directed paths to every
other node in the graph. Since the follower robots are not aware of
the destination, they have to stay connected with the informed robot
either directly or indirectly through concatenated paths, such that
the knowledge of the destination can be delivered to all the nodes
through the connected network. Hence, to complete the desired tasks,
maintaining connectivity of the underlying graph is necessary. }

\textcolor{black}{Collision avoidance among robots has not been considered
for rendezvous problems in existing literature (e.g., \cite{Lin2007,Lin2007a,Cort'es2006,dimarogonas2007,Su2010a,Gustavi2010}),
since it conflicts with the objective of meeting at a common goal.
To enable collision avoidance in this work, the workspace $\mathcal{F}$
is divided into a collision-free region $\Omega_{c}$ and a rendezvous
region $\Omega_{r}$, such that $\Omega_{c}\cup\Omega_{r}=\mathcal{F}$.
The rendezvous region $\Omega_{r}$ is a bounded disk area with radius
$R_{r}$ centered at the common destination $p^{\ast}$, while the
remaining area in $\mathcal{F}$ is the collision-free region $\Omega_{c}$.
Assume that the workspace $\mathcal{F}$ and the rendezvous region
$\Omega_{r}$ satisfy that $R_{w}\gg R_{r}$ and $R_{r}>R\left(N-1\right)$.
The classical rendezvous problem enables the robots to rendezvous
at $p^{\ast}$ with a desired orientation $\theta^{\ast}$ in $\Omega_{r}$,
we additionally constraint this model by requiring collision avoidance
among robots outside the neighborhood of common $p^{\ast}$ (i.e.,
$\Omega_{c}$). The main contribution of this work is to derive a
set of distributed controllers using only local information (i.e.,
position feedback from immediate neighbors and the absolute orientation
measurement) to perform rendezvous, ensure network connectivity, and
avoid collisions. To achieve these goals, the following assumptions
are required in the subsequent development.}
\begin{assum}
\textcolor{black}{\label{Ass1}The initial graph $\mathcal{G}\left(0\right)$
has a directed spanning tree with the informed node as the root.}
\end{assum}

\begin{assum}
\textcolor{black}{\label{Ass2}The destination $p^{\ast}$ and desired
orientation $\theta^{\ast}$ are achievable, which implies that $p^{\ast}$
and $\theta^{\ast}$ do not coincide with some unstable equilibria
(i.e., saddle points).}
\end{assum}

\section{\textcolor{black}{Control Design\label{sec:controller}}}

\subsection{\textcolor{black}{Dipolar Navigation Function}}

\textcolor{black}{Artificial potential field-based methods that use
attractive and repulsive potentials have been widely used to control
multi-robot systems. Due to the existence of local minima when attractive
and repulsive force are combined, robots can be trapped by local minima
and are not guaranteed to reach the global minimum of the potential
field. A navigation function is a particular category of potential
functions where the potential field does not have local minima and
the negative gradient vector field of the potential field guarantees
almost global convergence to a desired destination, along with (guaranteed)
collision avoidance, if the initial conditions do not lie within the
sets of measure zero. Formally, a navigation function is defined as
follows:}
\begin{defn}
\textcolor{black}{\label{Def1}\cite{Rimon_1990}\cite{Rimon1992}
Let $\mathcal{F}$ be a compact connected analytic manifold with a
boundary $\partial\mathcal{F}$, and $q_{d}$ be a goal point in the
interior of $\mathcal{F}$. A mapping $\varphi:\mathcal{F\rightarrow}\left[0,1\right]$
is a Navigation Function, if it satisfies the following conditions:
1) smooth on $\mathcal{F}$ (at least a $\mathcal{C}^{2}$ function);
2) admissible on $\mathcal{F}$, (uniformly maximal on the manifold
boundary $\partial\mathcal{F}$ and constraint boundary); 3) polar
on $\mathcal{F},$ ($q_{d}$ is a unique minimum); and 4) a Morse
function, (critical points of the navigation function are non-degenerate).}
\end{defn}
\textcolor{black}{The second condition in Definition \ref{Def1} establishes
that the generated trajectories are collision-free, since the resulting
vector field is transverse to the boundary of $\mathcal{F}$. The
third point indicates that, using a polar function on a compact connected
manifold with a boundary, all initial conditions are either brought
to a saddle point or to the unique minimum $q_{d}$. The requirement
that the navigation function is a Morse function ensures that the
initial conditions that bring the system to saddle points are sets
of measure zero \cite{Rimon_1990}. Given this property, all initial
conditions not within sets of measure zero are brought to the unique
minimum.}

\begin{figure}
\textcolor{black}{\centering{}\includegraphics[scale=0.35]{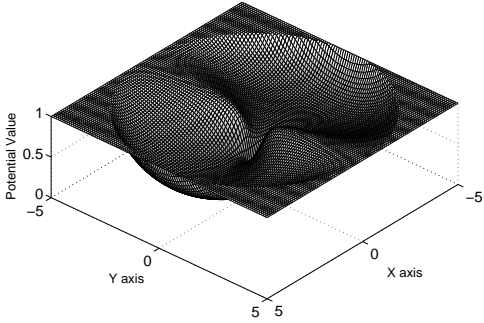}}

\textcolor{black}{\caption{An example of a dipolar navigation function with a workspace of $R_{w}=5$
and destination located at the origin with a desired orientation $\theta^{\ast}=0$.}
}

\textcolor{black}{\label{fig:dipolar}}
\end{figure}

\textcolor{black}{The navigation function introduced in \cite{Rimon_1990}
and \cite{Rimon1992} ensures global convergence of the closed-loop
system; however, the approach is not suitable for nonholonomic systems,
since the feedback law generated from the gradient of the navigation
function can lead to undesirable behaviors, which may be overcome
by extending the original navigation function to a Dipolar Navigation
Function in \cite{Tanner2000} and \cite{Tanner2003}. The flow lines
created in the potential field resemble a dipole, so that the flow
lines are all tangent to the desired orientation at the origin and
utilized by the vehicle to achieve the desired orientation. An example
of the dipolar navigation is shown in Fig. \ref{fig:dipolar}, where
the potential field has a unique minimum at the destination (i.e.,
$p^{\ast}=\left[0,0\right]^{T}$ and $\theta^{\ast}=0$)$,$ and achieves
the maxima at the workspace boundary of $R_{w}=5$. Note that the
surface $x=0$ divides the workspace into two parts and forces all
the flow lines to approach the destination parallel to the $y$-axis.}

\textcolor{black}{}

\textcolor{black}{Inspired by the work in \cite{Tanner2000} and \cite{Tanner2003},
the control strategy here is to develop a dipolar navigation function
for the informed robot, which creates a feasible nonholonomic trajectory
for the nonholonomic robot and guarantees the achievement of the specified
destination with a desired orientation, while other follower robots
aim to achieve consensus with the informed robot and maintain network
connectivity by using only local interaction with neighboring robots.
Following this idea, the dipolar navigation function is designed for
the informed node $i\in\mathcal{V}_{L}$ as $\varphi_{i}^{d}\left(t\right):\mathcal{F}\rightarrow[0,1],$
\begin{equation}
\varphi_{i}^{d}=\frac{\gamma_{d}}{\left(\gamma_{d}^{\alpha}+H_{d}\cdot\beta_{d}\right)^{1/\alpha}},\label{phi_d}
\end{equation}
where $\alpha\in\mathbb{R}^{+}$ is a tuning parameter. The goal function
$\gamma_{d}\left(t\right):\mathbb{R}^{2}\rightarrow\mathbb{R}^{+}$
in (\ref{phi_d}) encodes the control objective of achieving the desired
destination, which is specified by the distance from $p_{i}\left(t\right)\in\mathbb{R}^{2}$
to the destination $p^{\ast}\in\mathbb{R}^{2},$ and is designed as
$\gamma_{d}=\left\Vert p_{i}\left(t\right)-p^{\ast}\right\Vert ^{2}$.
The factor $H_{d}\left(t\right)\in\mathbb{R}^{+}$ in (\ref{phi_d})
creates a repulsive potential to align the trajectory of node $i$
at the destination with the desired orientation. The repulsive potential
factor is designed as
\begin{equation}
H_{d}=\varepsilon_{nh}+\left(\left(p_{i}-p^{\ast}\right)^{T}\cdot n_{d}\right)^{2},\label{dipolar}
\end{equation}
where $\varepsilon_{nh}$ is a small positive constant, and $n_{d}=\left[\begin{array}{cc}
\cos\left(\theta^{\ast}\right) & \sin\left(\theta^{\ast}\right)\end{array}\right]^{T}\in\mathbb{R}^{2}.$ A small disk area with radius $\delta_{1}<R$ centered at node $i$
is denoted as a collision region. To prevent a potential collision
between node $i$ and the workspace boundary, the function $\beta_{d}:\mathbb{R}^{2}\rightarrow\left[0,1\right]$
in (\ref{phi_d}) is designed as
\begin{equation}
\beta_{d}=\frac{1}{1+e^{-\frac{2}{\delta_{1}}\log\left(\frac{1-\epsilon}{\epsilon}\right)\left(d_{i0}-\frac{1}{2}\delta_{1}\right)}}\label{eq:B_i0}
\end{equation}
where $0<\epsilon\ll1$ is a positive constant, and $d_{i0}\triangleq R_{w}-\left\Vert p_{i}\right\Vert \in\mathbb{R}$
is the relative distance of node $i$ to the workspace boundary.}

\textcolor{black}{Since $\gamma_{d}$ and $\beta_{d}$ in (\ref{phi_d})
are guaranteed to not be zero simultaneously by Assumption \ref{Ass2},
the navigation function candidate in (\ref{phi_d}) achieves its minimum
of $0$ when $\gamma_{d}=0$ and its maximum of $1$ when $\beta_{d}=0$.
Our previous work in \cite{Kan.Dani.ea2012} proves that the original
navigation function with the form of $\varphi_{i}=\frac{\gamma_{i}}{\left(\gamma_{i}^{\alpha}+\beta_{i}\right)^{1/\alpha}}$
is a qualified navigation function. It is also shown in \cite{loizou2008}
that the navigation properties are not affected by the modification
to a dipolar navigation with the design of (\ref{dipolar}), as long
as the workspace is bounded, $H_{d}$ in (\ref{phi_d}) can be bounded
in the workspace, and $\varepsilon_{nh}$ is a small positive constant.
As a result, the decentralized navigation function $\varphi_{i}^{d}$
proposed in (\ref{phi_d}) can be proven to be a qualified navigation
function by following a similar procedure in \cite{loizou2008} and
\cite{Kan.Dani.ea2012}. From the properties of the navigation function,
it is known that almost all initial positions (except for a set of
measure zero points) asymptotically approach the desired destination.}

\textcolor{black}{To achieve consensus with the informed node while
ensuring network connectivity and collision avoidance, a local interaction
rule is designed for each follower node $i\in\mathcal{V}_{F}$ as
$\varphi_{i}^{f}\left(t\right):\mathcal{F}\rightarrow[0,1],$
\begin{equation}
\varphi_{i}^{f}=\frac{\gamma_{i}}{\left(\gamma_{i}^{\alpha}+\beta_{i}\right)^{1/\alpha}},\label{phi_f}
\end{equation}
where $\alpha\in\mathbb{R}^{+}$ is a tuning parameter. The goal function
$\gamma_{i}\left(t\right):\mathbb{R}^{2}\rightarrow\mathbb{R}^{+}$
in (\ref{phi_f}) encodes the control objective of achieving consensus
on the position between node $i$ and neighboring nodes $j\in\mathcal{N}_{i}$,
which is designed as 
\begin{equation}
\gamma_{i}=\sum_{j\in\mathcal{N}_{i}}\left\Vert p_{i}\left(t\right)-p_{j}\left(t\right)\right\Vert ^{2}.\label{goal fcn}
\end{equation}
Assume each node $i$ has a collision region defined as a small disk
with radius $\delta_{1}<R$, and an escape region defined as the outer
ring of the sensing area centered at the node with radius $r,$ $R-\delta_{2}<r<R,$
where $\delta_{2}\in\mathbb{R}^{+}$ is a predetermined buffer distance.
Any node $j\in$ $\mathcal{N}_{i}$ inside the collision region has
the potential to collide with node $i$, and each edge formed by node
$i$ and $j\in\mathcal{N}_{i}$ in the escape region has the potential
to break connectivity. To ensure collision avoidance and network connectivity,
the constraint function $\beta_{i}:\mathbb{R}^{2N}\rightarrow\left[0,1\right]$
in (\ref{phi_f}) is designed as 
\begin{equation}
\beta_{i}=\prod\nolimits _{j\in\mathcal{N}_{i}}b_{ij}B_{ij},\label{beta}
\end{equation}
by only accounting for nodes within its sensing area. Particularly,
$b_{ij}\left(p_{i},\mbox{ }p_{j}\right):\mathbb{R}^{2}\rightarrow\left[0,1\right]$
in (\ref{beta}) is a continuously differentiable sigmoid function,
designed as 
\begin{equation}
b_{ij}=\frac{1}{1+e^{-\frac{2}{\delta_{2}}\log\left(\frac{1-\epsilon}{\epsilon}\right)\left(R-\frac{1}{2}\delta_{2}-d_{ij}\right)}},\label{eq:b_ij}
\end{equation}
where $0<\epsilon\ll1$ is a positive constant. The designed $b_{ij}$
ensures connectivity of nodes $i$ and its neighboring nodes $j\in\mathcal{N}_{i}$
(i.e., nodes $j\in\mathcal{N}_{i}$ will never leave the sensing and
communication zone of node $i$ if node $j$ is initially connected
to node $i$). }

\textcolor{black}{Since collision avoidance among robots are only
required in $\Omega_{c}$, $B_{ij}(p_{i},\mbox{ }p_{j}):\mathbb{R}^{2}\rightarrow[0,1]$
in (\ref{beta}) is designed as
\begin{equation}
B_{ij}=\frac{1}{1+e^{-\frac{2}{\delta_{1}}\log\left(\frac{1-\epsilon}{\epsilon}\right)\left(d_{ij}-\frac{1}{2}\delta_{1}\right)}}\label{eq:B_ik}
\end{equation}
which indicates that collision avoidance is activated if the robots
are in $\Omega_{c}$, i.e., node $i$ is repulsed from other nodes
to prevent a collision in $\Omega_{c}$. If the robots are in $\Omega_{r}$,
the collision avoidance is deactivated by removing $B_{ij}$ from
$\beta_{i}$ in (\ref{beta}). Since $\Omega_{r}$ is defined by the
distance to the destination and only the leader in the group is informed
about the destination, the collision avoidance scheme designed in
(\ref{eq:B_ik}) is deactivated only when the leader is close enough
to the destination in $\Omega_{r}$}%
\footnote{\textcolor{black}{The network will be proven to be connected all the
time in the subsequent analysis, which implies that the distance between
any two nodes is bounded by $R\left(N-1\right)$. To ensure all followers
are in $\Omega_{r}$ when the collision avoidance scheme is deactivated,
the leader is required to deactivate the collision avoidance when
its distance to the destination is less than $R_{r}-R\left(N-1\right)$.}%
}\textcolor{black}{. }

\begin{figure}
\textcolor{black}{\centering{}\includegraphics[scale=0.4]{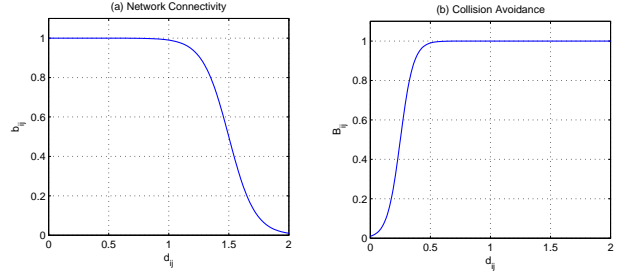}}

\textcolor{black}{\caption{The plot of $b_{ij}$ and $B_{ij}$ with $\delta_{1}=0.5$, $\delta_{2}=1$,
$R=2$, and $\epsilon=0.01$.}
}

\textcolor{black}{\label{fig:weight}}
\end{figure}

\textcolor{black}{That is $\beta_{i}$ in (\ref{phi_f}) for $\forall i\in\mathcal{V}_{F}$
switches from $\beta_{i}=\prod\nolimits _{j\in\mathcal{N}_{i}}b_{ij}B_{ij}$
to $\beta_{i}=\prod\nolimits _{j\in\mathcal{N}_{i}}b_{ij}$ if the
leader is close enough to the destination in $\Omega_{r}$, so that
collision avoidance among robots is not considered any more. The $b_{ij}$
and $B_{ij}$ are illustrated in Fig. \ref{fig:weight}. Note that
the constraint function in (\ref{beta}) is designed to vanish whenever
node $i$ intersects with one of the constraints in the environment,
(i.e., if node $i$ touches another node in $\Omega_{c}$, or separates
from adjacent nodes $j\in\mathcal{N}_{i}$ by distance of $R_{c}$).
Since $\gamma_{i}$ and $\beta_{i}$ in (\ref{phi_f}) will not be
zero simultaneously from their definitions, it is clear that $\varphi_{i}^{f}$
achieves its minimum of $0$ if $\gamma_{i}=0$ (i.e., consensus is
reached between node $i$ and its immediate neighbors), and $\varphi_{i}^{f}$
achieves its maximum of $1$ if $\beta_{i}=0$ (i.e., either the network
connectivity or collision constraint is met).}

\textcolor{black}{}

\subsection{\textcolor{black}{Control Development}}

\textcolor{black}{For brevity, $\varphi_{i}$ is used to represent
the potential function designed for each node $i$, where particularly
$\varphi_{i}=\varphi_{i}^{d}$ in (\ref{phi_d}) if $i\in\mathcal{V}_{L}$,
and $\varphi_{i}=\varphi_{i}^{f}$ in (\ref{phi_f}) if $i\in\mathcal{V}_{F}.$
The desired orientation for any robot $i\in\mathcal{V}$, denoted
by $\theta_{di}\left(t\right),$ is defined as a function of the negative
gradient of the decentralized function $\varphi_{i}$ as,
\begin{equation}
\theta_{di}\triangleq\arctan2\left(\begin{array}{cc}
-\frac{\partial\varphi_{i}}{\partial y_{i}}, & -\frac{\partial\varphi_{i}}{\partial x_{i}}\end{array}\right),\label{theta_d}
\end{equation}
where the mapping $\arctan2\left(\cdot\right):\mathbb{R}^{2}\rightarrow\mathbb{R}$
denotes the four quadrant inverse tangent function, and $\theta_{di}\left(t\right)$
is confined to the region of $\left(-\pi,\pi\right]$. By defining
$\theta_{di}\left\vert _{p^{\ast}}\right.=\arctan2\left(0,0\right)=\theta_{i}\left\vert _{p^{\ast}}\right.$,
$\theta_{di}$ remains continuous along any approaching direction
to the goal position. Based on the definition of $\theta_{di}$ in
(\ref{theta_d})
\begin{equation}
\nabla_{i}\varphi_{i}=-\left\Vert \nabla_{i}\varphi_{i}\right\Vert \left[\begin{array}{cc}
\cos\left(\theta_{di}\right) & \sin\left(\theta_{di}\right)\end{array}\right]^{T},\label{gradient}
\end{equation}
where $\nabla_{i}\varphi_{i}=\left[\begin{array}{cc}
\frac{\partial\varphi_{i}}{\partial x_{i}} & \frac{\partial\varphi_{i}}{\partial y_{i}}\end{array}\right]^{T}$ denotes the partial derivative of $\varphi_{i}$ with respect to
$p_{i}$, and $\left\Vert \nabla_{i}\varphi_{i}\right\Vert $ denotes
the Euclidean norm of $\nabla_{i}\varphi_{i}$. The difference between
the current orientation and the desired orientation for robot $i$
at each time instant is defined as 
\begin{equation}
\tilde{\theta}_{i}\left(t\right)=\theta_{i}\left(t\right)-\theta_{di}\left(t\right),\label{error}
\end{equation}
where $\theta_{di}\left(t\right)$ is generated from the decentralized
navigation function $\varphi_{i}$ and (\ref{theta_d}). Based on
the open-loop system in (\ref{eq:dynamics}), the controller for each
robot (i.e., the linear and angular velocity of robot $i$) is designed
as
\begin{equation}
v_{i}=k_{v,i}\left\Vert \nabla_{i}\varphi_{i}\right\Vert \cos\tilde{\theta}_{i},\label{v_linear}
\end{equation}
\begin{equation}
\omega_{i}=-k_{w,i}\tilde{\theta}_{i}+\dot{\theta}_{di},\label{v_angular}
\end{equation}
where $k_{v,i}$, $k_{w,i}$$\in\mathbb{R}^{+}$ denote the control
gains for robot $i$. The term $\dot{\theta}_{di}$ in (\ref{v_angular})
is determined as 
\begin{equation}
\dot{\theta}_{di}=k_{v,i}\cos(\tilde{\theta}_{i})\left[\begin{array}{c}
\sin\left(\theta_{di}\right)\\
-\cos\left(\theta_{di}\right)
\end{array}\right]^{T}\nabla_{i}^{2}\varphi_{i}\left[\begin{array}{c}
\cos\left(\theta_{i}\right)\\
\sin\left(\theta_{i}\right)
\end{array}\right],\label{thetad_dot}
\end{equation}
where $\nabla_{i}^{2}\varphi_{i}$ denotes the Hessian matrix of $\varphi_{i}$
with respect to $p_{i}$. Note that the computation of $\nabla_{i}\varphi_{i}$,
$\tilde{\theta}_{i}$, and $\dot{\theta}_{di}$ only requires local
position feedback and does not depend on communication with any neighbors,
which highlights the decentralized nature of the controllers in (\ref{v_linear})
and (\ref{v_angular}). Although the switch of $\beta_{i}$ will result
in a discontinuity of the controller in (\ref{v_linear}) and (\ref{v_angular})
when the leader enters $\Omega_{r}$ from $\Omega_{c}$, the controller
remains continuous within $\Omega_{r}$ and $\Omega_{c}$ respectively.
Substituting (\ref{v_linear}) into (\ref{eq:dynamics}), the closed-loop
system for robot $i$ can be obtained as
\begin{equation}
\dot{p}_{i}=\left[\begin{array}{c}
\dot{x}_{i}\\
\dot{y}_{i}
\end{array}\right]=k_{v,i}\left\Vert \nabla_{i}\varphi_{i}\right\Vert \cos\tilde{\theta}_{i}\left[\begin{array}{c}
\cos\theta_{i}\\
\sin\theta_{i}
\end{array}\right].\label{xy_close}
\end{equation}
After using the fact that $\left[\begin{array}{cc}
\cos\theta_{i} & \sin\theta_{i}\end{array}\right]\nabla_{i}\varphi_{i}=-\left\Vert \nabla_{i}\varphi_{i}\right\Vert \cos\tilde{\theta}_{i}$ from (\ref{gradient}), the closed-loop error systems can be expressed
as
\begin{equation}
\dot{p}_{i}=-k_{v,i}\nabla_{i}\varphi_{i},\text{ }i\in\mathcal{V}\text{.}\label{p_close}
\end{equation}
}

\section{\textcolor{black}{Connectivity and Convergence Analysis\label{sec:analysis}}}

\subsection{\textcolor{black}{Connectivity Analysis}}
\begin{thm}
\textcolor{black}{\label{thm:connectivity}The controller in (\ref{v_linear})
and (\ref{v_angular}) ensures that the initially connected spanning
tree structure is preserved when performing rendezvous for nodes with
kinematics given by (\ref{eq:dynamics}), as well as collision avoidance
among robots in $\Omega_{c}$.}\end{thm}
\begin{pf}
\textcolor{black}{The spanning tree structure in Assumption \ref{Ass1}
ensures that there exists a path from the informed node to every follower
node in $\mathcal{G}\left(0\right)$. To show every existing edge
in the directed spanning tree in $\mathcal{G}\left(0\right)$ is preserved,
consider a follower $i\in\mathcal{V}_{F}$ located at a position $p_{0}\in\mathcal{F}$
that causes $\beta_{i}=0$, which will be true when either only one
node $j$ is about to disconnect from node $i$ or when multiple nodes
are about to disconnect with node $i$ simultaneously. If $\beta_{i}=0$,
the navigation function $\varphi_{i}$ designed in (\ref{phi_f})
will achieve its maximum value. Driven by the negative gradient of
$\varphi_{i}$ in (\ref{p_close}), no open set of initial conditions
can be attracted to the maxima of the navigation function \cite{Rimon1992}.
Therefore, every edge in $\mathcal{G}$ is maintained and the directed
spanning tree structure is preserved for all time.}

\textcolor{black}{Similar to the proof of the preservation of each
link, if two nodes $i$ and $j$ are about to collide in $\Omega_{c}$,
that is $B_{ij}(p_{i},\mbox{ }p_{j})=0$ from (\ref{eq:B_ik}), then
the potential function $\varphi_{i}$ in (\ref{phi_f}) will reach
its maximum. Based on the properties of a navigation function driven
by the vector field in (\ref{p_close}), the system will not achieve
its maximum. Hence, collision among nodes is avoided. \qed}
\end{pf}

\subsection{\textcolor{black}{Convergence Analysis}}
\begin{lem}
\textcolor{black}{\label{Lem1}\cite{moreau2004,Moreau2005} Let }\textup{\textcolor{black}{$\mathcal{G}$}}\textcolor{black}{{}
be a directed graph of order $n$ and $L\in R^{n\times n}$ be the
associated (non-symmetric) Laplacian matrix. Consider a linear system
$\dot{x}\left(t\right)=-L\left(t\right)x\left(t\right),$ where $x\left(t\right)=\left[x_{1},\ldots,x_{n}\right]^{T}\in$$R^{n}$.
If the time-varying matrix $L\left(t\right)\in R^{n\times n}$ is
a piecewise continuous function of time with bounded elements, and
}\textup{\textcolor{black}{$\mathcal{G}$}}\textcolor{black}{{} has
a directed spanning tree for all $t\geq0$, then consensus is exponentially
achieved, i.e., $x_{1}=\cdots=x_{n}$.}\end{lem}
\begin{thm}
\textcolor{black}{Provided that }\textup{\textcolor{black}{$\mathcal{G}$}}\textcolor{black}{{}
has a spanning tree with the informed node as the root, the controller
in (\ref{v_linear}) and (\ref{v_angular}) ensures that all robots
with kinematics given by (\ref{eq:dynamics}) converge to a common
point with a desired orientation, in the sense that $\left\Vert p_{i}\left(t\right)-p^{\ast}\right\Vert \rightarrow0$
and $\left\vert \tilde{\theta}_{i}\left(t\right)\right\vert \rightarrow0$
as $t\rightarrow\infty$ $\forall i\in V$.}\end{thm}
\begin{pf}
\textcolor{black}{For the follower robots $i\in\mathcal{V}_{F}$,
the term $\nabla_{i}\varphi_{i}$ in (\ref{p_close}) is computed
from (\ref{phi_f}) as
\begin{equation}
\nabla_{i}\varphi_{i}=\frac{\alpha\beta_{i}\nabla_{i}\gamma_{i}-\gamma_{i}\nabla_{i}\beta_{i}}{\alpha(\gamma_{i}^{\alpha}+\beta_{i})^{\frac{1}{\alpha}+1}},\label{gradient2}
\end{equation}
where $\nabla_{i}\gamma_{i}$ and $\nabla_{i}\beta_{i}$ are bounded
in the workspace $\mathcal{F}$ from (\ref{goal fcn}) and (\ref{beta}),
and $\nabla_{i}\gamma_{i}$ and $\nabla_{i}\beta_{i}$ in (\ref{gradient2})
can be determined as 
\begin{equation}
\nabla_{i}\gamma_{i}=2\sum\nolimits _{j\in\mathcal{N}_{i}}\left(p_{i}-p_{j}\right),\label{eq:gama_gd}
\end{equation}
and 
\begin{equation}
\nabla_{i}\beta_{i}=2\sum\nolimits _{j\in\mathcal{N}_{i}}\left(\frac{\partial b_{ij}}{\partial d_{ij}}\right)\tfrac{\bar{b}_{ij}}{\left\Vert p_{i}-p_{j}\right\Vert }\left(p_{i}-p_{j}\right),\label{eq:beta_gd}
\end{equation}
respectively, where $\bar{b}_{ij}\triangleq\prod\nolimits _{l\in\mathcal{N}_{i},l\neq j}b_{il}.$
In (\ref{eq:beta_gd}), $\frac{\partial b_{ij}}{\partial d_{ij}}$
is 
\begin{equation}
\frac{\partial b_{ij}}{\partial d_{ij}}=-\tfrac{\frac{2}{\delta_{2}}\log\left(\frac{1-\epsilon}{\epsilon}\right)e^{-\frac{2}{\delta_{2}}\log\left(\frac{1-\epsilon}{\epsilon}\right)\left(R-\frac{1}{2}\delta_{2}-d_{ij}\right)}}{\left(1+e^{-\frac{2}{\delta_{2}}\log\left(\frac{1-\epsilon}{\epsilon}\right)\left(R-\frac{1}{2}\delta_{2}-d_{ij}\right)}\right)^{2}},\label{eq:bij_dij}
\end{equation}
which is negative, since $\delta_{2}$, $\frac{2}{\delta_{2}}\log\left(\frac{1-\epsilon}{\epsilon}\right)$,
and $e^{-\frac{2}{\delta_{2}}\log\left(\frac{1-\epsilon}{\epsilon}\right)\left(R-\frac{1}{2}\delta_{2}-d_{ij}\right)}$
are all positive terms. Substituting (\ref{eq:gama_gd}) and (\ref{eq:beta_gd})
into (\ref{gradient2}), $\nabla_{i}\varphi_{i}$ is rewritten as
\begin{equation}
\nabla_{i}\varphi_{i}=\sum\nolimits _{j\in\mathcal{N}_{i}}m_{ij}\left(p_{i}-p_{j}\right),\label{eq:phi_gd2}
\end{equation}
where 
\begin{equation}
m_{ij}=\frac{2\alpha\beta_{i}-2\left(\frac{\partial b_{ij}}{\partial d_{ij}}\right)\tfrac{\bar{b}_{ij}}{\left\Vert p_{i}-p_{j}\right\Vert }\gamma_{i}}{\alpha(\gamma_{i}^{\alpha}+\beta_{i})^{\frac{1}{\alpha}+1}}\label{eq:m_ij}
\end{equation}
is non-negative, based on the definitions of $\gamma_{i}$, $\beta_{i}$,
$\alpha$, $\bar{b}_{ij}$, and $\frac{\partial b_{ij}}{\partial d_{ij}}$
in (\ref{eq:bij_dij}). Using (\ref{p_close}) and (\ref{eq:phi_gd2})
yields the closed-loop system for each node $i$ as: 
\begin{equation}
\left\{ \begin{array}{cc}
\dot{p}_{i}(t)=-k_{v,i}\nabla_{i}\varphi_{i}^{d}, & i\in\mathcal{V}_{L}\\
\dot{p}_{i}(t)=-\sum\limits _{j\in\mathcal{N}_{i}}k_{v,i}m_{ij}\left(p_{i}-p_{j}\right), & i\in\mathcal{V}_{F},
\end{array}\right.\label{eq:pf2_1}
\end{equation}
which can be rewritten in a compact form as
\begin{equation}
\mathbf{\dot{p}}\left(t\right)=-\left(\pi\left(t\right)\otimes I_{2}\right)\mathbf{p}\left(t\right)+\mathbf{F}_{d},\label{eq:pf2_2}
\end{equation}
where $\mathbf{p}\left(t\right)=\left[\begin{array}{ccc}
p_{1}^{T}, & \cdots, & p_{N}^{T}\end{array}\right]^{T}\in\mathbb{R}^{2N}$ denotes the stacked vector of $p_{i},$ $\mathbf{F}_{d}=\left[\begin{array}{cccc}
-k_{v,i}\nabla_{i}^{T}\varphi_{i}^{d}, & 0, & \cdots, & 0\end{array}\right]^{T}\in\mathbb{R}^{2N}$ for $i=1$ (recall that $\mathcal{V}_{L}=\left\{ 1\right\} $ from
Section \ref{sec:problem}), $I_{2}$ is a $2\times2$ identity matrix,
and the elements of $\pi\left(t\right)\in\mathbb{R}^{N\times N}$
are defined as
\begin{equation}
\pi_{ik}\left(t\right)=\left\{ \begin{array}{cc}
\sum\nolimits _{j\in\mathcal{N}_{i}}k_{v,i}m_{ij}, & i=k\\
-k_{v,i}m_{ik}, & k\in\mathcal{N}_{i},i\neq k\\
0, & k\notin\mathcal{N}_{i},i\neq k.
\end{array}\right.\label{eq:pf2_3}
\end{equation}
Using the fact that $m_{ij}$ is non-negative from (\ref{eq:m_ij}),
and $k_{v,i}$ is a positive constant gain in (\ref{v_linear}), the
off-diagonal elements of $\pi\left(t\right)$ are negative or zero,
and its row sums are zero. Hence, $\pi\left(t\right)$ is a Laplacian
matrix. Since the informed node acts as the root in the spanning tree
structure in $\mathcal{G}$, the first row of $\pi\left(t\right)$
is composed of all zeros, which indicates that the motion of the informed
node is not dependent upon the motion of the followers. From Lemma
\ref{Lem1} and the properties of the dipolar navigation function
in (\ref{phi_d}), the first term in (\ref{eq:pf2_2}) indicates that
$p_{1}=\cdots=p_{N},$ and the second term implies that $p_{1}\rightarrow p^{\ast},$
and hence, $p_{i}\rightarrow p^{\ast}$ $\forall$ $i\in\mathcal{V}$. }

\textcolor{black}{Note that the properties of the developed dipolar
navigation function in (\ref{phi_d}) ensure that the informed node
achieves the specified destination with the desired orientation. If
the informed node always tracks its desired orientation $\theta_{di}$
and all the followers move along with the informed node, the group
will achieve the destination with desired orientation. To show that
$\left\vert \tilde{\theta}_{i}\right\vert \rightarrow0$ , we take
the time derivative of $\tilde{\theta}_{i}\left(t\right)$ in (\ref{error})
and use (\ref{eq:dynamics}) to develop the open-loop orientation
tracking error system as $\overset{\cdot}{\tilde{\theta}}_{i}=\omega_{i}-\dot{\theta}_{di}$.
Using (\ref{v_angular}), the closed-loop orientation tracking error
is 
\begin{equation}
\overset{\cdot}{\tilde{\theta}}_{i}=-k_{w}\tilde{\theta}_{i},\label{theta_close}
\end{equation}
which has the exponentially decaying solution $\tilde{\theta}_{i}\left(t\right)=\tilde{\theta}_{i}\left(0\right)e^{-k_{w}t}$.
\qed}\end{pf}
\begin{rem}
\textcolor{black}{Since the network is proven to be connected in Theorem
\ref{thm:connectivity}, the distance between any two nodes is bounded
by $R\left(N-1\right)$, where $R$ is the sensing radius and $N$
is the number of nodes. By requiring the informed robot to move within
a region with a distance to the workspace boundary greater than $R\left(N-1\right)$,
the followers are guaranteed to avoid collision with the workspace
boundary by achieving consensus with the informed robot.}
\end{rem}

\begin{rem}
\textcolor{black}{The previous analysis is based on the simplification
that only one informed node is considered. The result can be generalized
to multiple informed nodes by using containment control theory. Containment
control is a particular class of consensus problems in which all nodes
are grouped into followers and leaders, and the followers, under the
influence of leaders through local information exchange, converge
to a desired region (i.e., a convex hull) formed by the leaders' states.
Some recent results are reported in \cite{Notarstefano2011,Cao2009,Mei2012}
for containment control. In our recent work in \cite{Kan2012b}, a
decentralized method is developed to influence followers in a social
network to reach a common desired state (i.e., within a convex hull
spanned by the leaders), while maintaining interaction among the followers
and leaders. As a special case of \cite{Kan2012b}, if each leader
is assigned the same destination, the convex hull formed by leaders
will shrink to the common destination, and the followers will converge
to this desired destination. Therefore, following a similar approach
in \cite{Kan2012b}, all nodes can be proven to converge to the common
destination, if multiple informed nodes are considered. }
\end{rem}

\begin{rem}
\textcolor{black}{The switch of the controller (\ref{v_linear}) and
(\ref{v_angular}) from $\Omega_{c}$ to $\Omega_{r}$ will not affect
the stability of the system. Theorem 3.2 in \cite{DeCarlo2000} states
that a switched nonlinear system is stable if the associated Lyapunov-like
function $V_{i}$ in each region $\Omega_{i}$ is nonincreasing, and
$V_{i}$ is also nonincreasing when switching occurs. It is proven
that $\sum_{i}^{N}\varphi_{i}$ is a qualified Lyapunov function in
\cite{Kan.Dani.ea2012}, and following a similar approach as \cite{Kan.Dani.ea2012},
$\sum_{i}^{N}\varphi_{i}$is nonincreasing in $\Omega_{c}$ and $\Omega_{r}$,
respectively. To show that the Lyapunov function $\sum_{i}^{N}\varphi_{i}$
is nonincreasing when switching occurs, note that the denominator
of $\varphi_{i}^{f}$ in (\ref{phi_f}) is nondecreasing when switching
from $\Omega_{c}$ to $\Omega_{r}$ due to the fact that $B_{ij}\in\left[0,1\right]$,
which results in a nonincreasing $\varphi_{i}^{f}$. By invoking Theorem
3.2 in \cite{DeCarlo2000}, the system remains stable when the switch
occurs from $\Omega_{c}$ to $\Omega_{r}$.}
\end{rem}

\section{\textcolor{black}{Simulation\label{sec:simulation}}}

\begin{figure}
\textcolor{black}{\centering{}\includegraphics[scale=0.8]{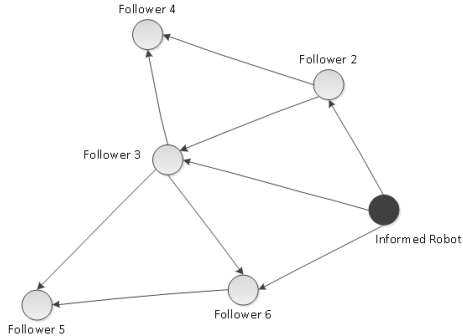}}

\textcolor{black}{\caption{Network topology with directed edges indicating the directed interaction
between two nodes.}
}

\textcolor{black}{\label{fig:graph}}
\end{figure}

\textcolor{black}{The following numerical simulation demonstrates
the performance of the controller developed in (\ref{v_linear}) and
(\ref{v_angular}) in a scenario in which a group of six mobile robots
with the kinematics in (\ref{eq:dynamics}) are navigated to the common
destination $p^{\ast}=\left[\begin{array}{cc}
0 & 0\end{array}\right]^{T}$ with the desired orientation $\theta^{\ast}=0.$ The limited communication
and sensing zone for each robot is assumed as $R=2$ m and $\delta_{1}=\delta_{2}=0.4$
m The tuning parameter $\alpha$ in (\ref{phi_d}) is selected to
be $\alpha=1.2$. The workspace $\mathcal{F}$ is a disk area centered
at the origin with radius $R_{w}=50,$ and the rendezvous region is
a disk area with radius $R_{r}=11.5$. The collison avoidance scheme
of the group is deactivated only when the informed robot is within
a distance less than 1.5 to the common destination. The group of mobile
robots is arbitrarily deployed in the workspace and forms a connected
graph. The interaction between nodes is described by the network topology
with a spanning tree as shown in Fig. \ref{fig:graph}, where the
light dots denote the follower robots and the dark dot denotes the
informed node. The informed node is randomly selected from the group,
and is the only node aware of the desired destination $p^{\ast}$
and orientation $\theta^{\ast}$. }

\begin{figure}
\textcolor{black}{\centering{}\includegraphics[scale=0.32]{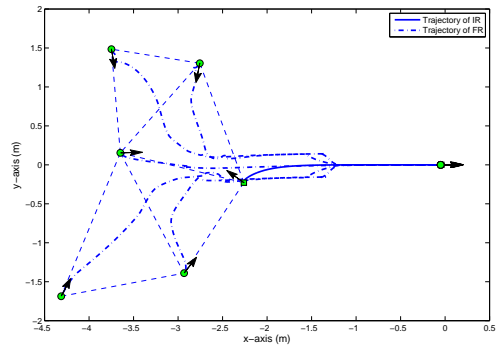}}

\textcolor{black}{\caption{Plot of robot trajectories with solid line and dot-dash line indicating
the trajectory of the informed robot (IR) and the follower robot (FR),
respectively.}
}

\textcolor{black}{\label{fig:trajectory}}
\end{figure}

\begin{figure}
\textcolor{black}{\centering{}\includegraphics[scale=0.32]{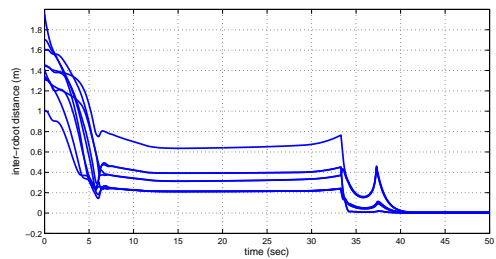}}

\textcolor{black}{\caption{The evolution of inter-robot distance.}
}

\textcolor{black}{\label{fig:dis}}
\end{figure}

\textcolor{black}{The control laws in (\ref{v_linear}) and (\ref{v_angular})
yield the simulation results shown in Fig. \ref{fig:trajectory} and
Fig. \ref{fig:dis}. Fig. \ref{fig:trajectory} shows the trajectory
for each robot, where the associated arrows indicate the initial or
final orientation. To demonstrate the collision avoidance among robots
and connectivity of existing links, the evolution of the inter-robot
distance is plotted in Fig. \ref{fig:dis}. As shown in Fig. \ref{fig:dis},
the inter-robot distance decreases significantly for the first few
seconds. Since the robots are moving in the collision free region
initially, where collision avoidance is activated, the inter-robot
distance stops to decrease when two robots are close to each other.
Once the robots enter the rendezvous region, where collision avoidance
is deactivated, the inter-robot distance decreases again to perform
the desired rendezvous. Note that inter-robot distance is maintained
less than the radius $R=2$ m, which indicates that connectivity of
the underlying graph is preserved. }

\textcolor{black}{}
\textcolor{black}{}
\textcolor{black}{}

\section{\textcolor{black}{Conclusion\label{sec:conclusion}}}

\textcolor{black}{A decentralized dipolar navigation function-based
time-varying controller is developed to navigate a network of mobile
robots to a common destination with a desired orientation while ensuring
network connectivity and collision avoidance, using only local sensing
information from one-hop neighbors. A distinguishing feature of the
developed decentralized approach is that no inter-agent communication
is required to complete the network consensus objective. Another distinguishing
feature is that the more general problem of directed networks is considered,
where only one robot is informed of the global objective while other
robots coordinate their motions to perform the cooperative task by
using local information feedback from immediate neighbors. Although
consensus among robots is ensured in this work, the rate of convergence
is not considered. Generally, the rate of convergence depends on the
network topology, which is a function of the roles of nodes (i.e.,
informed nodes or followers) and their interactions. A different set
of informed nodes may lead to different convergence rates. Additional
methods could be developed to optimize performance metrics such as
the degree of connectivity and the convergence rate of the network
in scenarios where the set of informed nodes can be determined and/or
positioned a prior.}

\bibliographystyle{IEEEtran}
\bibliography{master,ncr,\string"E:/nonliner control/my paper/bibtex/bib/ncrbibs/master\string",\string"E:/nonliner control/my paper/bibtex/bib/ncrbibs/ncr\string",\string"C:/ZHEN KAN/NCR/My Paper/bibtex/bib/ncrbibs/master\string",\string"C:/ZHEN KAN/NCR/My Paper/bibtex/bib/ncrbibs/ncr\string"}

\end{document}